\documentclass[referee]{aa}
\usepackage{graphics}
\usepackage{psfig}

\begin{document}

\title{Proper Motions and CCD-photometry of Stars in the Region of the Open Cluster NGC$\,$1513}
\author{V.N. Frolov\inst{1} \and E.G. Jilinski \inst{1,2}
\and J.K. Ananjevskaja \inst{1} \and E.V. Poljakov\inst{1}
\and N.M. Bronnikova \inst{1} \and D.L. Gorshanov \inst{1} }

   \offprints{V.N.Frolov, The Main Astronomical Observatory,
Pulkovo, St.\,Petersburg, Russia. email: vfrol@gao.spb.ru}

   \institute{The Main Astronomical Observatory, Pulkovo, St.\,Petersburg,
Russia
  \and  Laborat\'orio Nacional de Computa\c c\~ ao Cient\' \i fica / MCT,
Petr\'opolis, RJ, Btazil}

\date{}

\authorrunning{Frolov}

\titlerunning{Open Cluster NGC$\,$1513}

\maketitle

%______________________________________________________________________________

\begin{abstract}
The results of astrometric and photometric investigations of the
poorly studied open cluster NGC$\,$1513 are presented.  The proper
motions of 333 stars with a root-mean-square error of
$1.9\;{\rm mas\,yr^{-1}}$ were obtained by means of  the automated
measuring complex "Fantasy". Eight astrometric  plates covering
the time interval of 101 years were measured and a total of 141
astrometric cluster members identified. $BV$ CCD-photometry was obtained
for stars in an area $17\arcmin\times 17\arcmin$
centered on the cluster. Altogether 33 stars with high reliability
were considered to be cluster members by two criteria. The estimated age
of  NGC\,1513 is $2.54\cdot 10^8\,$ years.\footnote{Tables~ 2 and 3
are only available in electronic form at the CDS via anonymous ftp to
cdsarc.u-strasbg.fr (130.79.128.5) or
via http://cdsweb.u-strasbg.fr/cgi-bin/qcat?J/A+A/}

\end{abstract}

%______________________________________________________________________________

\section{Introduction}

  The open cluster NGC\,1513 is located in the Perseus
  constellation. Its equatorial and galactic coordinates are:

$\alpha=4^{\rm h} 09^{\rm m} 98^{\rm s}$,  $\delta=+49\degr 31\arcmin$;
$\ell=152\fdg 6$, $b=-1\fdg 57$      $ (2000.0)$.\\
In the Tr\"umpler(1930)         %(\cite{trumpler})
catalogue it is classified as II\,2m a moderately
rich cluster with little central concentration.
It was investigated
astrometrically  by Bronnikova(1958a)           %(\cite{bronnikovaa})
who  determined the proper motions (PM) using a single pair
of plates with an epoch difference of 55 years.
Barhatova and Drjahlushina(1960)    %(\cite{barhatovaa})
published photographic and photovisual magnitudes of 49 stars from
Bronnikova's (1958b)               %(\cite{bronnikovab})
list. Del Rio and Huestamendia(1988)  %(\cite{delrio})
(RH) obtained photoelectric $UBV$ magnitudes of 31 stars and
photographic $RGU$ magnitudes for 116 stars in the cluster region.
Spectra and radial velocities of stars in this area have not been
investigated. As the epoch difference has increased significantly
since the time of the last and only attempt to investigate
proper motions of the stars in the region of the open cluster NGC\,1513
we decided to study it using the material available at present.

%______________________________________________________________________________

\section{Astrometry}

%______________________________________________________________________________

   The observational material  belongs to the Normal
   astrograph collection of the Pulkovo observatory and dates
   from 1899 to 2000 (Table~ 1).
 The cluster region from the USNO-A 2.0 catalogue was used as an
additional plate (epoch 1954.8). The scale of the Normal
astrograph plates is ${\rm 60\arcsec\,mm^{-1}}$.
The positions of 383 stars in an area
$60\arcmin\times 60\arcmin$ centered on the  cluster NGC\,1513
were measured.
  For the first time our plates were scanned by means of the
automated measuring complex "Fantasy". As there is no description
of the complex in English we present it in the Appendix.

\begin{table}
      \caption[]{Astrometric plates}
\begin{tabular}{lcll}
\hline
 Plate         &   Exposure      & \,\,\,Epoch     &        Quality\\
          &    (min)     &         &        \\
\hline
\multicolumn{4}{c}{Early  epoch}\\
A  371  &         unknown &              1899  Nov  30   &      good\\
D  451  &              16 &              1951  Feb   7   &      high\\
  5011  &              26 &              1954  Nov  25   &      good\\
  5017  &              26 &              1954  Nov  30   &      high\\
\multicolumn{4}{c}{Second  epoch}\\
 18152  &              20 &              1999  Nov  19   &      good\\
 18159  &              25 &              1999  Dec  12   &      poor\\
 18170  &              30 &              2000  Feb   2   &      poor\\
 18177  &              25 &              2000  Feb   6   &      good\\
\hline
\end{tabular}
\end{table}

    The proper motions were determined by the line-method
(the star positions tranferred to one system and arranged in
chronological order of the epochs of observation were compared).
The details of the method are presented in the paper of
Jilinski et al.(2000).                                  %(\cite{jilinski})
The plate 5017 with the faintest images was used as
the central plate. It was orientated using 64 stars in common with
stars in the "TYCHO" catalogue stars so that the $Y$-axis
was parallel to the sky meridian. The reference stars for the
reduction were selected by several approximations.
 Each step consists of two successive parts.
At the first step all the stars of the region were considered as
reference stars. For each star graphs were considered where
the epochs of the plates were plotted as abscissa and the coordinate
$X$ or $Y$ of the stars as ordinates. Two straight lines were drawn by a
least square fitting. Their inclinations correspond to the proper motion
components $\mu _x$ and $\mu _y$ of the star. After this preliminary
determination of PM the first rough selection of cluster members was made.
Those of the stars that  were located on the vector point diagram (VPD)
in a circle with a radius of three standard deviation
from the center of the concentration of points corresponding to
the  cluster were considered to be members of the cluster.
As a result of the first approximation a new list of reference stars
was compiled according to the generally accepted rule: that they must
be uniformly distributed in the studied region excluding the cluster area,
have small proper motions and belong to a chosen range of $B$ magnitudes.
As the $UBV$ magnitude of only a few stars were determined in
the investigated area we had to use the considerably less precise values
of $B$ magnitudes of the USNO-A 2.0 catalogue.
Thus 39 stars in the interval between $13.5$ and $14.5$ mag were selected.
The linear method of reduction was used.
 From the second approximation the selection of cluster members
was made according to Sanders method(1971)             %(\cite{sanders})
assuming that the proper motion distributions of the cluster and
field stars on the VPD were bivariate  Gaussian distributions,
circular for the cluster $\Phi_c (\mu_x,\mu_y)$ and elliptical for
the field $\Phi_f (\mu_x,\mu_y)$.
Modelling of the resulting  distribution was fulfilled by the method
of maximum likelihood. The values of the parameters of the corresponding
system of the non-linear equations were determined by  computer iterations.
All the formulas are given in Jilinski et al(2000).
  The individual membership probability of a star was determined
by its position on VPD and calculated by the following formula:

$$ P \left(\mu_x,\mu_y\right) = \frac{N_c\Phi_c \left(\mu_x,\mu_y\right)}
{N_c\Phi_c \left(\mu_x,\mu_y\right) + N_f\Phi_f \left(\mu_x,\mu_y\right)} \eqno
(3)$$

     $N_c$ - normalized number of cluster stars,

     $N_f$ - normalized number of field stars,

$\mu_{x_i}$, $\mu_{y_i}$ - proper motion in $x$ and $y$ for the $\mbox{\it
i}^{th}$ star.

In order to have a fast fit in the solution of the system
parameters close to the real initial
values of parameters were adopted. In the final calculation of
probabilities the following values were used:
the relative number of cluster members and field stars $N_c/{N_f}=0.5$;\\
centers of the distributions on the VPD:                                \\
the field  $\mu_{x_0} = 1.34$ ,    $\mu_{y_0}=  0.45$ ;
the cluster $\mu_{x_0}= -0.9$,   $\mu_{y_0}= 0.44$ ; \\
standard deviations:   the field $\Sigma_{x}= 7.0$, $\Sigma_{y}= 5.0$ ;
                       the cluster $\sigma_{x,y}= 1.5$.\\
 All the values are in ${\rm mas\,yr^{-1}}$.
 Each iteration ended with plotting the histogram of
probabilities. Its analysis permitted an estimation of the minimal
value of probability for a star to be considered a cluster member.
The final histogram is shown in Fig.~1.

% FIG 1 *******

\begin{figure} [t]
\centering{
\vbox{\psfig{figure=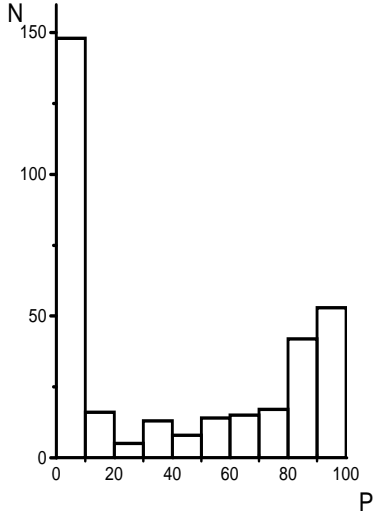,width=6cm,height=8cm}}\par}
\caption []{The final histogram of star membership probabilities.}
\end{figure}

  The magnitude equation (ME) or the dependence of
the  determined proper motions on star magnitudes was investigated at
each step of the membership selection. During the first iteration the ME was
derived from an investigation of the VPD for assumed field stars
with proper motion components corrected for the motion
of the Sun towards the solar apex. The values of these components were
calculated with the help of the tables published by
Zhukov(1966).                                 %(\cite{zhukov})
Then ME was estimated by the selected
cluster members anew in each iteration.  The direction and
inclination of the ME
(for $X$-axis:   $-1.06 \pm 0.29\, {\rm mas\,yr^{-1}}$,
 for $Y$-axis:   $-0.46 \pm 0.36\, {\rm mas\,yr^{-1}}$)
were derived from the field stars in the whole
interval of $B$ magnitudes from $10.5$ to $15$ mag.
These values were used to correct those of the original
catalogue. After the selection of cluster members
the estimated ME was : $X$:  $-1.95 \pm 0.42\, {\rm mas\,yr^{-1}}$,
$Y$:  $-0.78 \pm 0.42\, {\rm mas\,yr^{-1}}$.
The final catalogue corrected according to these values appeared
to be free of ME within the limits of the accuracy of the determination
of the PM of the stars. The final VPD is given in Fig.~2.

% FIG 2 *******

\begin{figure} [t]
\centering{
\vbox{\psfig{figure=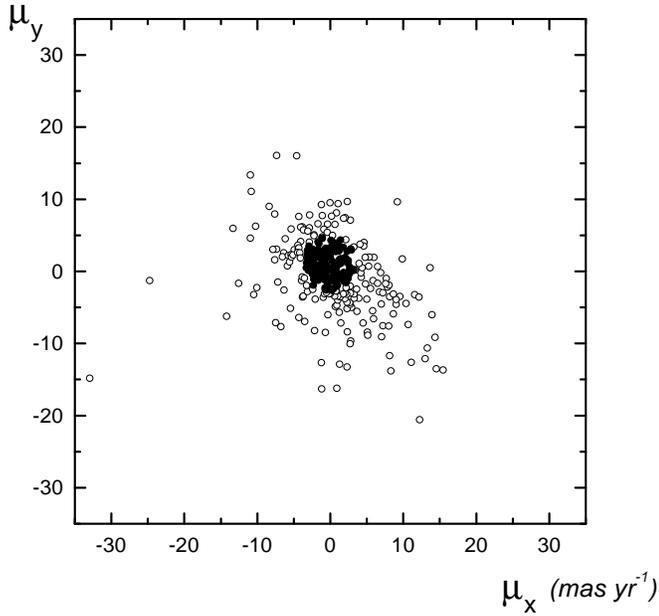,width=9cm,height=9cm}}\par}
\caption []{The final VPD of NGC\,1513. }
\end{figure}

   The rms errors of the proper motion components of the stars are:
$$ \sigma _x = \pm 1.8\;{\rm mas\,yr^{-1}},\;\, \sigma _y =\pm 1.9\;{\rm
mas\,yr^{-1}}.  $$

The astrometric results are given in Table~ 2. Columns:
  1) - the number of a star according to our study (the reference
stars are marked by the symbol *),
 2),3) - the rectangular coordinates of a star relative to the star N 197
 (N 115 RH) in $arcmin$,
 4) - $B$ magnitudes from the USNO -A 2.0,
 5)- the number of plates,
 6),7),8),9) - PM $\mu_{x}$,$\mu_{y}$ and the corresponding standard
 errors in ${\rm mas\,yr^{-1}}$,
 10) - membership probabilities in\,$\%$,
 11) - "bl" marks blends, "bs" - objects whose coordinates
 were determined with large errors. It also contains the numbers of
the stars in the BD, HD and Hipparcos catalogues.

Due to various reasons (blending, wrong identification, etc.)
we were not able to determine the PM of 50 stars from the total number
of 383 stars. Astrometric  cluster members were considered to be
stars with a membership probability $P\ge 50\%$.
  The total number of such stars is 141 (the probability of 112 of
them $P\ge 70\%$).
%______________________________________________________________________________

\section{Photometry}

% FIG 3 *******
\begin{figure*} [t]
\centering{
\vbox{\psfig{figure=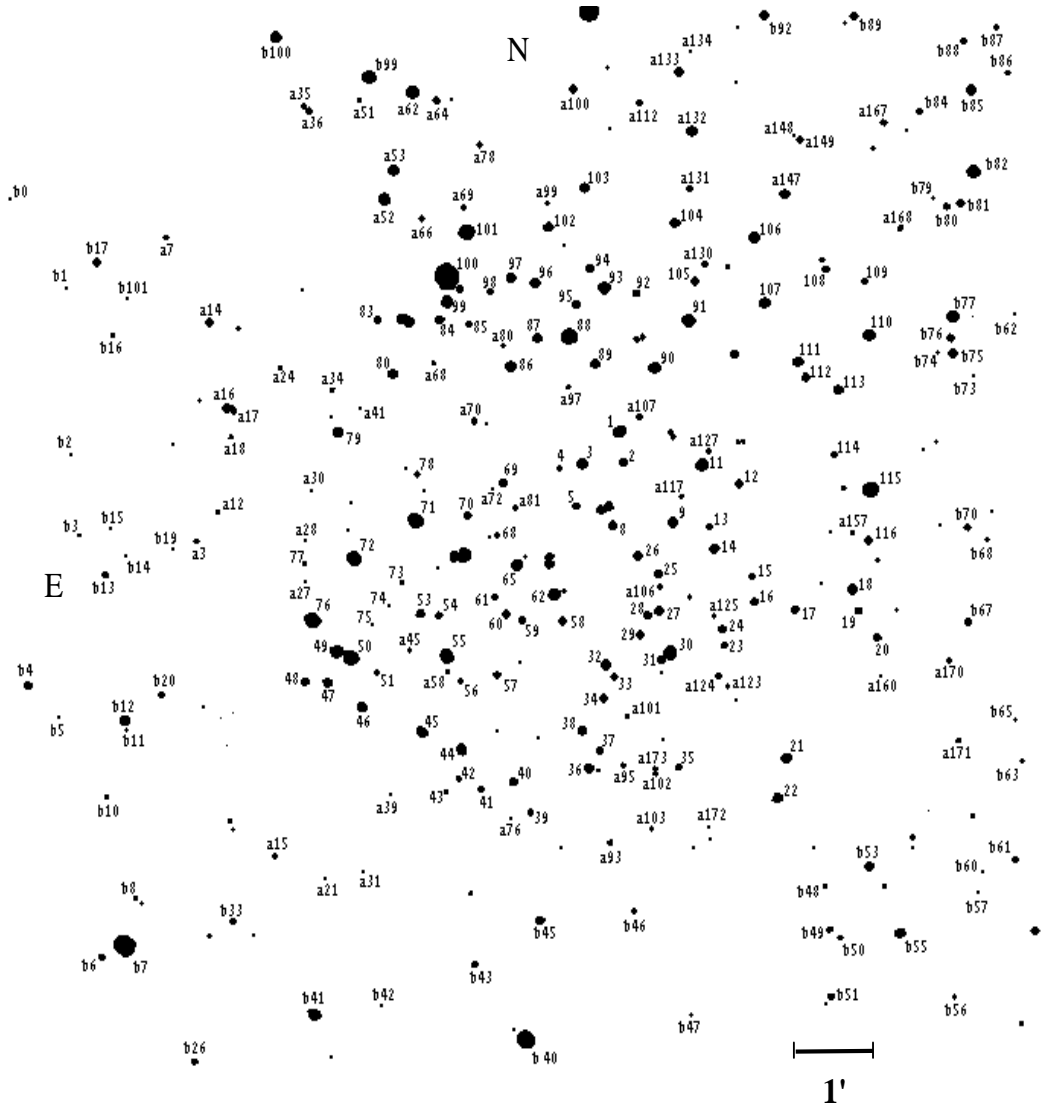,width=15cm,height=16cm}}\par}
\caption []{CCD map of the central part of the cluster NGC\,1513.
The numbers of stars are given according to the photometric study:
numbers without an index -RH; stars marked by the index 'a' were not
measured by RH; stars marked by index 'b' are beyond the RH map.}
\end{figure*}

  The $B$ and $V$ magnitudes of stars in the central part of
the investigated area were obtained with the 320 mm mirror astrograph ZA-320
of the Pulkovo Observatory equipped with the CCD-receiver ST-6 with a
fixed TC241 matrix. A detailed description of the telescope and its
resources is given in the paper of Bekyashev et al.(1998).  %(\cite{bekyashev})
The area of the field determined by the scale(${65\arcsec\,mm^{-1}}$)
and the dimensions of the matrix is equal to $9.3\arcmin\times 7\arcmin$.
Due to the small area of the field the observations of $B$ and $V$
colours were limited to the area $17\arcmin\times 17\arcmin$
where all stars from the RH paper are located.
The $UBV$ magnitudes  of 31 stars from  Table~ 2 of their paper were used
as photoelectric standards.
Besides stars marked by RH numbers there are in the same
field  a considerable amount of stars without any numbers. They
are marked by numbers beginning with "$a$". Some stars that are
beyond the limits of the RH map were also measured and are marked by
"$b$". The map of the cluster is given in Fig.~3 and the CCD
magnitudes and colours - in Table~ 3 with the columns:
 1) - the RH number of a star,
 2),9),14) - the star number according to our study,
 3),10),15) - CCD photometry in the $V$ band,
 4),11),16) - CCD colours ($B-V$),
 5),6)- photoelectric(RH) $V$ and $(B-V)$,
 7),12) and 17) - notes.

% FIG 4 *******

\begin{figure} [t]
\centering{
\vbox{\psfig{figure=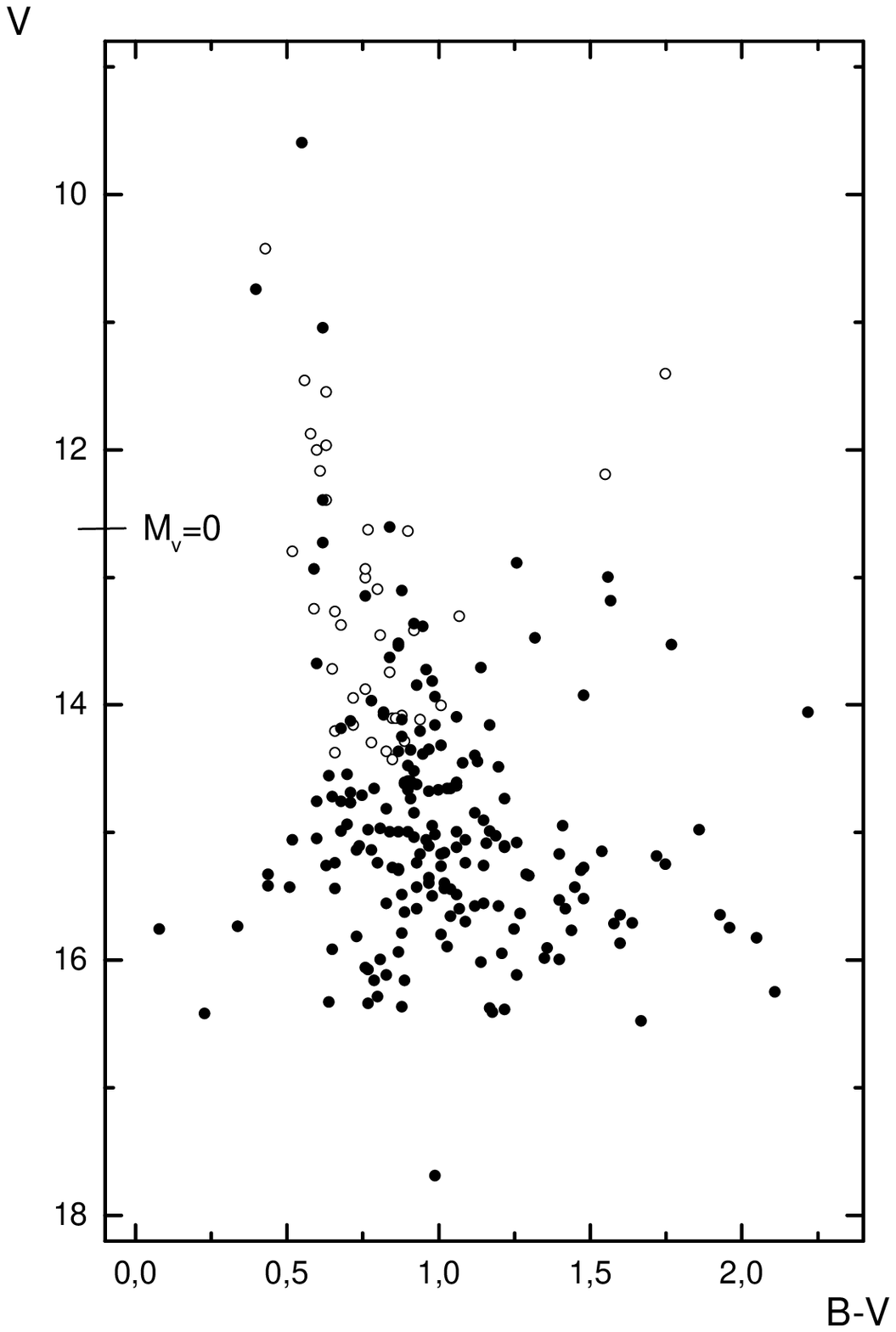,width=8.5cm,height=14cm}}\par}
\caption []{CMD diagramm of the cluster NGC\,1513. $\bullet$ - stars with
the membership probability $P\geq 50\%$.}
\end{figure}

The external rms errors calculated for the standard
stars with zero mean deviations is $\pm 0.012\;{\rm mag}$
in the $B$ pass-band  and $\pm 0.010\;{\rm mag}$ in $V$. After
the reduction of the instrumental values to the standards no residual
colour equestion was detected. The internal  rms errors increase from the bright
to the faint stars. Their mean values in the interval from 9.6 mag
to 16.37 mag are $\sigma_B = \pm 0.07\;{\rm mag}, \sigma_V = \pm 0.05\;{\rm mag}$.
%______________________________________________________________________________

\section{Analysis and conclusion}

\setcounter{table}{3}
\begin{table*}  %Table 4
\caption{Members and probable cluster members.}
\begin{tabular}{rrccrrccrcc}
\hline

\multicolumn{5}{c}{Members} & \multicolumn{6}{c}{Probable members}   \\
$N_{pm}$&$N_{phot}$&$M_V$&$(B-V)_0$& $P$& $ N_{phot}$& $M_V$&$(B-V)_0$& $N_{phot}$&$M_V$&$(B-V)_0$ \\
  124&  b53&   1.27&  0.09&  59&        4&    2.63&  0.26&     a 69&    3.55&  0.12  \\
  126&   37&   1.68&  0.22&  94&       12&    2.30&  0.48&     a 70&    2.79&  0.35  \\
  127&   38&   0.85&  0.14&  95&       15&    2.34&  0.31&     a 78&    3.19&  0.34  \\
  130&   45&   0.40&  0.09&  83&       16&    1.99&  0.23&     a 80&    3.30&  0.69  \\
  132&  b12&   0.33&  0.09&  88&       17&    2.05&  0.12&     a 81&    2.95&  0.48  \\
  133&   b7&  -2.18& -0.24&  95&       20&    2.05&  0.36&     a 93&    3.15&  0.58  \\
  159&   48&   1.11& -0.02&  90&       23&    2.56&  0.27&     a 97&    3.38&  0.68  \\
  160&   76&  -1.15& -0.11&  66&       24&    1.74&  0.30&     a112&    2.56&  0.34  \\
  161&   47&   0.64& -0.08&  94&       28&    1.87&  0.23&     a117&    3.47&  0.10  \\
  162&   49&   0.19& -0.15&  77&       29&    2.07&  0.30&     a123&    3.51&  0.59  \\
  163&   50&  -0.60& -0.07&  94&       33&    2.42&  0.52&     a124&    2.82&  0.26  \\
  164&   72&   0.54& -0.06&  93&       34&    2.03&  0.39&     a127&    3.05&  0.37  \\
  166&   53&   1.34&  0.05&  93&       35&    2.39&  0.39&     a131&    2.95&  0.16  \\
  167&   54&   1.77& -0.01&  80&       39&    2.38&  0.01&     a133&    2.05&  0.37  \\
  168&   71&  -1.06& -0.04&  86&       41&    2.53&  0.11&     a157&    3.39&  0.73  \\
  169&   55&  -0.64& -0.04&  91&       42&    2.88&  0.39&     a160&    3.29&  0.36  \\
  173&   69&   1.60& -0.01&  68&       43&    2.99&  0.40&     a167&    2.66&  0.34  \\
  174&   58&   1.50&  0.19&  94&       51&    2.15& -0.07&     a168&    2.97&  0.45  \\
  175&   62&  -0.41&  0.88&  93&       52&    2.68&  0.20&     a170&    2.79&  0.30  \\
  178&    5&   1.76&  0.18&  84&       56&    2.89&  0.31&     a171&    2.50&  0.30  \\
  184&    2&   1.55&  0.05&  66&       57&    2.08&  0.04&     a172&    2.39&  0.20  \\
  185&    1&  -0.21& -0.04&  62&       59&    2.16&  0.04&     b  0&    2.83& -0.01  \\
  190&    9&   0.77&  0.01&  94&       61&    2.21&  0.16&     b  1&    2.43&  0.25  \\
  194&   18&   1.48&  0.21&  91&       70&    1.94&  0.03&     b  2&    2.33&  0.03  \\
  215&  110&   0.49&  0.13&  82&       73&    2.55&  0.35&     b  5&    2.83&  0.35  \\
  216&  113&   1.51&  0.27&  90&       74&    2.44& -0.07&     b 10&    2.48&  0.49  \\
  217&  111&   1.50&  0.18&  89&       75&    2.75&  0.30&     b 13&    1.58&  0.01  \\
  219&  106&   1.14&  0.17&  56&       77&    2.69&  0.20&     b 16&    2.36&  0.14  \\
  222&   90&   0.66& -0.01&  93&       83&    2.51&  0.39&     b 17&    1.47&  0.15  \\
  229&   88&  -1.20&  1.08&  90&       84&    2.65& -0.04&     b 19&    2.99&  0.26  \\
  232&   97&   1.82&  0.18&  81&       85&    2.50&  0.07&     b 20&    2.45&  0.42  \\
  234&  101&  -0.73& -0.09&  52&       89&    1.52&  0.04&     b 33&    2.51&  0.55  \\
  239&   80&   1.69&  0.11&  56&       92&    2.39&  0.23&     b 42&    2.84&  0.37  \\
     &     &       &      &    &      105&    2.41&  0.32&     b 43&    2.00&  0.39  \\
     &     &       &      &    &      108&    1.99&  0.24&     b 46&    3.09&  0.42  \\
     &     &       &      &    &      109&    2.67&  0.18&     b 48&    3.39&  0.14  \\
     &     &       &      &    &      112&    2.37&  0.10&     b 49&    2.39&  0.17  \\
     &     &       &      &    &      114&    2.24&  0.25&     b 50&    3.18&  0.21  \\
     &     &       &      &    &     a  7&    2.02&  0.26&     b 51&    2.24&  0.45  \\
     &     &       &      &    &     a 12&    2.06&  0.23&     b 56&    3.21&  0.06  \\
     &     &       &      &    &     a 14&    2.13&  0.24&     b 57&    3.77&  0.50  \\
     &     &       &      &    &     a 18&    2.11& -0.02&     b 60&    3.41&  0.47  \\
     &     &       &      &    &     a 21&    3.78&  0.55&     b 65&    3.33&  0.20  \\
     &     &       &      &    &     a 24&    3.51&  0.16&     b 67&    1.95& -0.03  \\
     &     &       &      &    &     a 27&    2.63& -0.01&     b 70&    2.50&  0.55  \\
     &     &       &      &    &     a 28&    3.02&  0.22&     b 74&    3.76&  0.21  \\
     &     &       &      &    &     a 30&    3.45&  0.09&     b 75&    1.91&  0.25  \\
     &     &       &      &    &     a 31&    2.88&  0.21&     b 76&    2.45&  0.29  \\
     &     &       &      &    &     a 39&    3.03&  0.60&     b 79&    3.80&  0.51  \\
     &     &       &      &    &     a 41&    3.68&  0.13&     b 80&    2.65&  0.48  \\
     &     &       &      &    &     a 45&    2.63&  0.13&     b 81&    2.06&  0.33  \\
     &     &       &      &    &     a 53&    2.00&  0.22&     b 89&    2.38&  0.50  \\
     &     &       &      &    &     a 58&    3.34&  0.54&     b 92&    2.01&  0.22  \\
     &     &       &      &    &     a 66&    3.55&  0.22&     b101&    2.97&  0.53  \\
     &     &       &      &    &     a 68&    3.73&  0.10&         &        &        \\

\hline
\end{tabular}
\end{table*}

% FIG 5 *******

\begin{figure} [t]
\centering{
\vbox{\psfig{figure=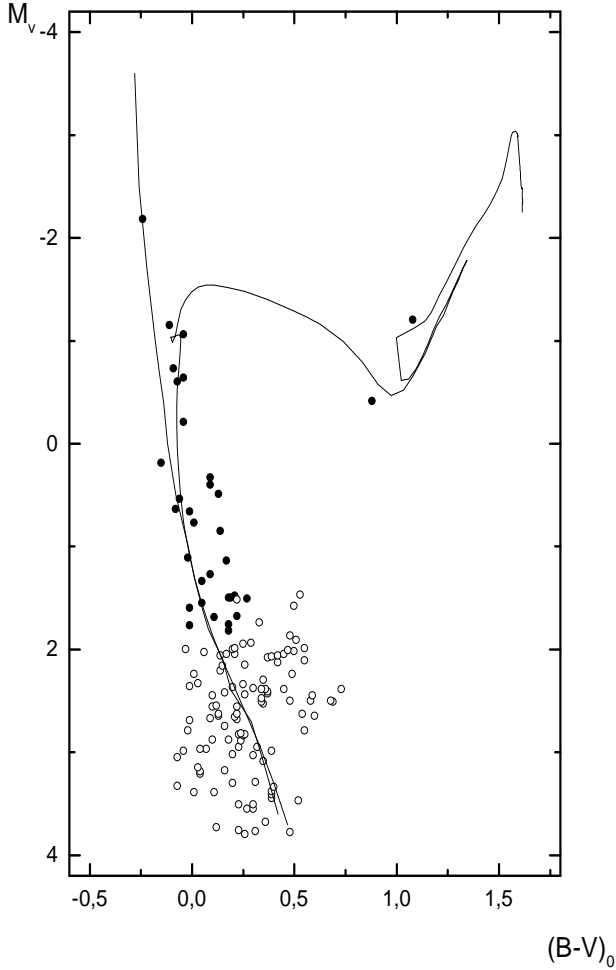,width=8.5cm,height=14cm}}\par}
\caption []{The $M_V\&(B-V)_0$ diagramm of all cluster members. ZAMS -
 Schmidt-Kaler(1965). Isochrone - $z=0.0019, age= 2.54\cdot 10^8 yr$.}
\end{figure}

 All the available $B,V$ - photometry  resulted in the $V$ \& $B-V$ (CMD)
diagram given in Fig.~4. Stars with membership probability $P\ge 50\%$
are plotted by solid circles. Obviously it is difficult to define
the Main sequence (MS) due to
contradictions in the published data on selective absorption in
the cluster region (estimates of the mean $E_{B-V}$ vary from 0.52 mag
to 0.70 mag). With the absence of U magnitudes of the stars we had to
use the ZAMS position from the RH paper. The value of the
colour-excess $E_{B-V}$ = 0.67 was calculated from the formula
$E_{B-V} = 0.72E_{G-R} -0.005(E_{B-V})^2$ (Steinlin,1968)    %(\cite{steinlin})
with $E_{G-R}=0.94$ mag and apparent distance modulus $(V-M_V)=12.61$
using the true modulus $(V-M_V)_0 = 10.60$.
Stars located on the main sequence within the limits $\pm 3\sigma$
are considered to be cluster members according to photometric criterion.
In Fig.~5 the  cluster members (according to  photometric
and astrometric criteria) are plotted as solid circles.
Open circles mark probable cluster members (only by the photometric
criterion) as their magnitudes are fainter than the limiting values of the
Normal astrograph old plates. In Fig.~5 there are
two stars with high membership probabilities at the late evolution
stages: N 175: $M_V= -0.41$\,mag, $(B-V)_0 = 0.88$\,mag, $P=93\%$ and
N 299 : $M_V=-1.20$\,mag, $(B-V)_0 = 1.08$\,mag, $P=90\%$;
ZAMS -  Schmidt-Kaler(1965).                    %(\cite{schmitkaler})
The members  and  probable  members are listed in  Table~ 4.

   The estimation of the age of the cluster is based on the set of
isochrones for stars with mean masses from 0.16 to 7 $M_\odot$
and different metallicities (Z) published  on the website of the Padove
research groupe\footnote{http://www. pleadi.astro.pd.it} and
described in the work of Girardi et al. (2000).  %(\cite{girardi})
The superposition of  various isochrones on the CMD shows an almost
ideal coincidence of the $2.54\cdot 10^8$ yr isochrone for $Z=0.019$
with the cluster diagram (Fig.~5). The loop that marks the ending
of hydrogen burning in the core, compression  of the core and
hydrogen burning in the thick layer is well occupied by stars.
The star No 175 is located almost at the base of the red giant branch,
and the star No 229 -at the stage of helium burning. Corresponding to
this isochrone the turn-off point of the MS is $(B-V)_0 = -0.07$ mag.

  It is also interesting to note  that the MS of NGC\,1513 on
the CMD in  absolute coordinates almost completely coinsides
with the MS of the nearby cluster NGC\,1528. Their ages also
coincide. The age and MS position of the latter were discussed in
the paper by Mermilliod(1981).   %(\cite{mermilliod})
Are these facts accidental? It is still
possible that Barhatova(1963)     %(\cite{barhatovab})
who treated them along with NGC\,1545
cluster as a triplet related by a common origin was right.
RH did not confirm the physical connection of NGC\,1513 and NGC\,1528
because of the essential difference in their distances. We think that
it is worthwhile to obtain data on the membership of stars of
NGC\,1528 not only by the photometric diagram but also by proper
motions of its stars. The Pulkovo observatory collection
contains a set of negatives taken with the Normal astrograph
from 1959 to the present time.

%______________________________________________________________________________

\section{Appendix}

   The automated measuring complex "Fantasy" is a programmed
precision microdensitometer with capabilities determined by
the software developed for it. The base of the machine is
a massive (1500 kg) metal table with a polished surface on
which a carriage on aerostatic bearings moves. The carriage is
actuated by two line electric motors and its position is determined by
a laser interferometer. The scanning system consists of a cathode ray
tube with a program control beam and photoelectronic multiplier with
the measured plate between them. The characteristics of the positioning
and scanning systems are given in Table~ 5.

 The algorithm of position measurements developed for "Fantasy"
is efficient for measurements in all ranges
of stellar images on a plate, from the brightest with a diameter
600-800 microns to very faint hardly
discernible images consisting of separate darkened grains. In reality
the secondary forms of the images are
measured. These forms are constructed during the processing of the
images by iterative operations:
filtration, contrasting and normalising. This permits the reduction
of a wide range of stellar images on a plate
to a set of compact objects of the regular form with centres defined
as the centres of their photometric
weights. A deleting operation was included in the algorithm for
measurements of old plates with a Gauthier
grid.
The developed algorithm for position measurements has high operational
and metrological  characteristics. It
operates accurately in conditions of considerable background density
differences, low quality images and in
the  presence of photographic blemishes and defects. The standard
error of measurement depends on the
quality of a photographic material and equals 0.14-0.60 microns
for the whole range of images.
\begin{table*}  %Table 5
\caption{"Fantasy" characteristics}
\begin{tabular}{ll}
\hline
Positioning system &  Scanning system \\
Carriage operating field $370\times370mm$ & View window   $4\times4 mm$ \\
Time of positioning  $4s$ &  Range of reading     $ 20kp\,s^{-1}$ \\
Accuracy of positioning  $ 1 micron$  & Aperture $ 2-3 micron$ \\
Accuracy of position  measuring $ 0.32 micron$ & Resolution  $1 micron^{2}$ \\
Speed of the carriage movement $330 mm\,s^{-1}$ & Dynamic range $2D$ \\
\hline
\end{tabular}
\end{table*}

\begin{acknowledgements}
        We would like to thank Dr. Z.Kadla for many useful comments
        on the manuscript and investigators of Pulkovo observatory
        M.Sidorov, A.Devyatkin, I.Grigorjeva and V.Kouprianov
        for their collaboration in the observations.
        E.G.Jilinski thanks FAPERJ for the financial support under the
        contract E-26/152.221/2000.

\end{acknowledgements}

%______________________________________________________________________________

{}

\end{document}